\documentclass[english]{sbrt}
\usepackage[english]{babel}
\usepackage[utf8]{inputenc}
\usepackage{tipa}
\usepackage{amsmath} 
\usepackage{booktabs} 
\usepackage{multirow} 
\usepackage[table]{xcolor}
\usepackage{makecell}
\usepackage{url}
\usepackage{tikz}
\usetikzlibrary{positioning, arrows.meta, shapes.geometric}
\usepackage{graphicx}

\begin{document}

\title{Extracting accent features in spoken Brazilian Portuguese without sociolinguistic labels}

\author{Pedro H. L. Leite, Pedro Benevenuto Valadares and Luiz W. P. Biscainho
\thanks{Pedro Leite, PEE/COPPE, UFRJ, Rio de Janeiro-RJ, e-mail: pedro.lopes@smt.ufrj.br; Pedro Valadares, Faculdade de Engenharia Elétrica e Computação (FEEC), UNICAMP, Campinas-SP, e-mail: p204483@dac.unicamp.br; Luiz Biscainho, DEL/Poli \& PEE/COPPE, UFRJ, Rio de Janeiro-RJ, wagner@smt.ufrj.br. This work was partially supported by the National Council for Scientific and Technological Development, CNPq (306395/2025-8).}%
}


\maketitle

\markboth{SUBMITTED TO XLIV BRAZILIAN SYMPOSIUM ON TELECOMMUNICATIONS AND SIGNAL PROCESSING - SBrT 2026, SEPTEMBER 29TH TO OCTOBER 2ND, 2026}{}

\begin{abstract}

Regional accent classification in Brazilian Portuguese (pt-BR) suffers from the need for reliable labeling. While large self-supervised learning (SSL) speech models are powerful, their training pipelines dilute sociophonetic information, since accent labels are generally not reliable or are not used in training objectives. This work introduces a novel workflow for feature extraction using only acoustic labels. By isolating explicit regional accent landmarks and using a phoneme-based forced aligner (ZIPA), our targeted feature set captures dialectal variance more effectively than utterance embeddings, demonstrating that localized features can outperform general-purpose architectures on accent-related tasks using minimal and objective data labels.
\end{abstract}

\section{Introduction}
As AI models mature and enable large-scale speech processing applications, regional accent classification remains a primary challenge, particularly for Brazilian Portuguese (pt-BR). While prior work dedicated efforts to modeling dialectal variation in this language~\cite{yneguti1999, batista2018deteccao, tostee2021classificacao, matos2023accent}, it has been shown that the classification of spontaneous speech with cross-domain adaptation still represents a gap, as specifically mentioned in~\cite{batista2018deteccao} and~\cite{matos2023accent}. These findings systematically highlight the challenges of dialect identification across multi-source datasets using neural architectures, showing that massively pretrained models using the self-supervised learning (SSL) paradigm, such as Wav2Vec 2.0~\cite{baevski2020wav2vec2}, HuBERT~\cite{hsu2021hubert}, and XLSR-53~\cite{xlsr2021unsupervised} or even deeply supervised such as ECAPA-TDNN~\cite{desplanques2020ecapa} are unable to solve the general task with current data availability. 

For speech, these models are usually fine-tuned for either automatic speech recognition (ASR)~\cite{baevski2020wav2vec2,xlsr2021unsupervised,hsu2021hubert} or speaker identity verification~\cite{desplanques2020ecapa}. In the case of ASR, since different phoneme realizations would generally be mapped to the same text characters, the inner layers of SSL models learn to actively suppress acoustic variations caused by channel noise, speaker identity, and, unfortunately, regional accents. This dilutes the localized dialectal information and confuses it within the high-dimensional embedding space, which hardens the task of accent-related feature extraction.


In large speech datasets, each speaker typically records utterances using a single accent. This structural configuration prevents natural disambiguation between speaker identity and regional accent, as standard corpora lack parallel multi-accent recordings from the same individual~\cite{zhou2026multiscale}. Consequently, when deep architectures are optimized for downstream tasks like speaker identification or verification, the resulting latent spaces tightly couple speaker timbre and dialectal features. This entanglement effect fails to reflect the paralinguistic reality that a single human speaker can theoretically navigate multiple regional accents while maintaining a consistent underlying vocal identity, as explored in~\cite{melechovsky2024dart}.

This work aims to show that this information entanglement makes detecting accents much harder for pretrained models fine-tuned for other downstream tasks, since adapting classifiers from these embedding spaces comes with the burden of removing all the noisy side information present in their high-dimensional latent space. Also, to avoid the disentanglement need, this paper demonstrates that regional accents can be identified with state-of-the-art accuracy using low-dimensional acoustic feature spaces and accent marker localization.

On the other hand, sociolinguistic labels are often misleading for accent classification~\cite{matos2023accent, callou1996variacao}, as multiple factors affect how people speak. Gathering information like birthplace, current residence, and family origin helps narrow down possibilities, but in practice, other diverse factors can affect people's accents and make these labels unreliable. In fact, simply listening to samples from accent-labeled Brazilian Portuguese datasets reveals many disparities.

Given this context, this work proposes

\begin{itemize}
    \item a purely audio-driven pipeline that uses multilingual forced aligners to isolate specific sociophonetic markers in spoken Brazilian Portuguese, applying classical signal processing to classify these markers without relying on sociolinguistic labels;
    
    \item an extensive analysis of the optimal features for detecting various accent markers in Brazilian Portuguese, including a feature-space search and a performance comparison against large pretrained models;
    
    \item the application of these features to regional accent detection, demonstrating feature explainability alongside speaker and cross-dataset consistency.
\end{itemize}

\section{Accent markers detection}

Previous works in linguistics~\cite{callou1996variacao} and signal processing~\cite{batista2018deteccao, tostee2021classificacao} highlight the importance of certain phoneme realizations as accent markers that can be used for regionality identification. In this work, speakers are specifically classified according to /s/ coda realizations, /r/ coda realizations, and the presence or absence of /d/ and /t/ palatalization. Since the accent markers must be first localized in written text, these class choices come from the linguistic simplicity of using syntax rules for localization. Also, these classes can already help to uniquely identify many of the regionalistic distinctions following the mapped distributions found in~\cite{cardoso2014atlas}.

The collection of detection features began with the manual annotation of data from multiple sources, using well-established pt-BR speech datasets: CORAA~\cite{candido2023coraa}, Mozilla Common Voice~\cite{ardila2020common}, ColingPB\footnote{https://repositorio.ufpb.br/jspui/handle/123456789/23184}, BRSpeechDF~\cite{filho-etal-2025-brspeech}, CML-TTS~\cite{oliveira2023cmltts}, Certas Palavras~\cite{araujo2026certas}, CETUC~\cite{alencar2008speech}, gneutralspeech male~\cite{leite2022corpus} and female~\cite{sbrt2023}. For ColingPB, speakers were extracted by diarization through PyAnnote\footnote{https://huggingface.co/pyannote/speaker-diarization-3.1}. Only speakers with consistent behavior (realization choice) were considered for annotation. 

For each phonological task, first the phoneme realization positions were localized in time using ZIPA~\cite{zhu2025zipa} in forced-alignment mode. Attempts to use Allosaurus~\cite{li2020universal} and CUPE~\cite{rehman2025cupe} models for this acoustic localization resulted in alignments and phoneme predictions that were not accurate enough for the proposed task.

ZIPA extracts IPA~\cite{ipa1999handbook} phoneme characters directly from audio, but cannot consistently disambiguate between different possible regional realizations. This is because the phonetic reference labels seen in ZIPA training were extracted from text-only Grapheme-to-Phoneme (G2P) systems Charsiu~\cite{zhu2022charsiug2p} and Epitran~\cite{mortensen2018epitran}, which do not contain regional accent variability and instead rely on global representations of Brazilian Portuguese. However, as ZIPA is trained on multilingual objectives, their internal layers encode the acoustic disambiguation information directly, as they are relevant for matching ground-truth phonemes found in other languages. Based on this observation, this work seeks to experimentally validate the hypothesis that  ZIPA logits can be easily adapted for this task.



To isolate target sociophonetic markers, the following heuristic pipeline is applied: define all valid Brazilian Portuguese IPA mappings for the /s/, /r/, /d/, and /t/ phonemes; transcribe each sentence using ZIPA with CTC to generate character-level timestamps; with phoneme candidates in hand, iterate through the predicted sequences, flagging /s/ and /r/ codas if the subsequent phone is a consonant or word boundary, and, similarly, flagging /d/ and /t/ targets if immediately followed by an /i/-related phone; extract the enunciation timestamps for all flagged targets and generate a 160ms audio clip centered on that timestamp, capturing the short-term acoustic context around the spike of interest.

\section{Feature Extraction}
  For each 160 ms clip, different groups of features are extracted for evaluation. The first group comprises six spectral moments computed over the power spectrum restricted to 500–8000 Hz ($N_{FFT} = 4096$, Hann window): spectral centroid ($M_1$), variance ($M_2$), skewness ($M_3$), excess kurtosis ($M_4$), peak frequency, and a log band-energy ratio $\log_{10}(E_{5\text{–}8\text{kHz}} / E_{2\text{–}4\text{kHz}})$. 

  The second group of features is an MFCC feature vector. For each clip, 13 Mel-frequency cepstral coefficients are computed using 32 ms analysis windows with 10 ms hops, a 40-channel Mel filterbank spanning 50–8000 Hz, and an FFT size of 512 points at 16 kHz. First- and second-order delta
  coefficients are appended, yielding a 39-dimensional vector per frame; the final clip representation is the average vector across all frames.

  The third group consists of ZIPA softmax probabilities extracted at $\pm$2 frames around the detected spike, corresponding to the phones most relevant to the contrast (e.g., \{s, \textipa{S}, \textipa{\textctc}, \textipa{\:s}, z, \textipa{Z}, f, x\} for /s/-coda; \{\textipa{R}, \textipa{\textinvscr}, \textipa{\textscr}, \textipa{\textcrh}, \textipa{X}, r, h, x\} for /r/-coda; \{t, d, j, \textipa{S}, \textipa{Z}, i\} for /d/-/t/ palatalization). For comparison, this third group of features is also extracted with Allosaurus. 
  
  The fourth group of features is comprised of extracted using SoTA deep learning models: XLSR-53~\cite{conneau2020unsupervised} and XLSR-53 fine-tuned for pt-BR\footnote{https://huggingface.co/jonatasgrosman/wav2vec2-large-xlsr-53-portuguese}, HuBERT~\cite{hsu2021hubert}, Wav2Vec 2.0~\cite{baevski2020wav2vec2}, Resemblyzer\footnote{https://github.com/resemble-ai/resemblyzer}, and ECAPA-TDNN~\cite{desplanques2020ecapa}.
  
  Speaker-level features are formed by averaging clip-level vectors over all detected instances for that speaker. 
  
\section{Classifying speaker realization choices}
\begin{figure}[htpb]
\centering
\resizebox{\columnwidth}{!}{%
\begin{tikzpicture}[
    every node/.style={align=center},
    marker/.style={rectangle, draw, rounded corners, fill=gray!15, thick, minimum width=2.4cm, minimum height=0.5cm},
    ipa/.style={ellipse, draw, fill=white, thick, inner sep=2pt},
    region/.style={rectangle, draw=none, font=\footnotesize\itshape, text width=2.2cm, inner sep=1pt}
]

\node[marker] (scoda) at (1.5, 0) {/s/ Coda};

\node[ipa] (chiado) at (0.2, -0.9) {[\textipa{S}]};
\node[region, below=0.03cm of chiado] (chiadoreg) {Rio de Janeiro,\\Florianópolis,\\Belém};

\node[ipa] (sibilant) at (2.8, -0.9) {[\textipa{s}]};
\node[region, below=0.03cm of sibilant] (sibilantreg) {São Paulo,\\South};

\node[marker] (dtpalat) at (7.0, 0) {/d/--/t/ + /i/};

\node[ipa] (palat) at (5.4, -0.9) {[\textipa{dZ\super{i}}]/[\textipa{tS\super{i}}]};
\node[region, below=0.03cm of palat] (palatreg) {Rio de Janeiro,\\São Paulo,\\Minas Gerais};

\node[ipa] (nonpalat) at (8.6, -0.9) {[\textipa{d\super{i}}]/[\textipa{t\super{i}}]};
\node[region, below=0.03cm of nonpalat] (nonpalatreg) {Northeast};

\node[marker] (rcoda) at (4.25, -3.1) {/r/ Coda};

\node[ipa] (carioca) at (1.7, -4.0) {[\textipa{x}]/[\textipa{h}]};
\node[region, below=0.03cm of carioca] (cariocareg) {Rio de Janeiro,\\Fluminense};

\node[ipa] (tap) at (4.25, -4.0) {[\textipa{R}]};
\node[region, below=0.03cm of tap] (tapreg) {São Paulo (City),\\Porto Alegre};

\node[ipa] (caipira) at (6.8, -4.0) {[\textipa{\textrtailr}]};
\node[region, below=0.03cm of caipira] (caipirareg) {Interior SP,\\Paraná,\\Goiás};

\draw[-Latex, thick] (scoda) -- (chiado) node[midway, left, font=\scriptsize\sffamily, xshift=-1pt, yshift=2pt] {Chiado};
\draw[-Latex, thick] (scoda) -- (sibilant) node[midway, right, font=\scriptsize\sffamily, xshift=1pt, yshift=2pt] {Sibilant};
\draw[dotted, thick] (chiado) -- (chiadoreg);
\draw[dotted, thick] (sibilant) -- (sibilantreg);

\draw[-Latex, thick] (dtpalat) -- (palat) node[midway, left, font=\scriptsize\sffamily, xshift=-1pt, yshift=2pt] {Palatalized};
\draw[-Latex, thick] (dtpalat) -- (nonpalat) node[midway, right, font=\scriptsize\sffamily, xshift=1pt, yshift=2pt] {Preserved};
\draw[dotted, thick] (palat) -- (palatreg);
\draw[dotted, thick] (nonpalat) -- (nonpalatreg);

\draw[-Latex, thick] (rcoda) -- (carioca) node[midway, left, font=\scriptsize\sffamily, xshift=-1pt, yshift=2pt] {Carioca};
\draw[-Latex, thick] (rcoda) -- (tap) node[midway, fill=white, inner sep=1pt, font=\scriptsize\sffamily] {Tap};
\draw[-Latex, thick] (rcoda) -- (caipira) node[midway, right, font=\scriptsize\sffamily, xshift=1pt, yshift=2pt] {Caipira};
\draw[dotted, thick] (carioca) -- (cariocareg);
\draw[dotted, thick] (tap) -- (tapreg);
\draw[dotted, thick] (caipira) -- (caipirareg);

\end{tikzpicture}%
}
\caption{Taxonomy of targeted Brazilian Portuguese accent phonetic markers. The targeted phonological variables (gray boxes) map to distinct phonetic realizations (circles), which are strong indicators of specific regional dialects (label texts).}
\label{fig:accent_tree}
\end{figure}
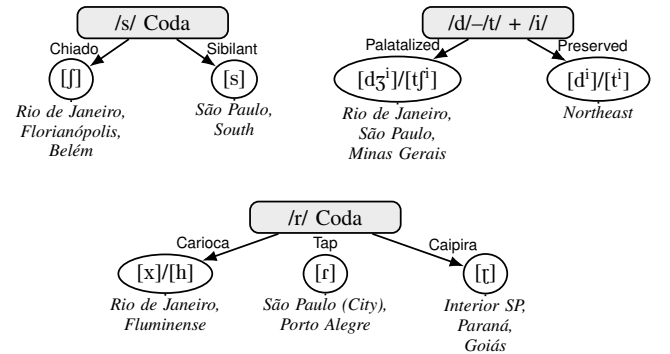

To understand the effectivity of the chosen features, a classifier is trained to match human labels on the regional choices for each speaker. The class choices reflect a simplification of the markers shown in ALiB maps~\cite{cardoso2014atlas} as distinctive features for Brazilian regional accents as shown in Figure~\ref{fig:accent_tree}. 

For /s/ coda, two classes are defined, based on the place of articulation of the fricative realization. The \textit{chiado} class corresponds to speakers who produce the palato-alveolar fricative \textipa{[S]} in coda position, a salient feature of the Rio de Janeiro city accent. The \textit{sibilant} class corresponds to the canonical unvoiced alveolar fricative \textipa{[s]}, the dominant realization in the majority of Brazilian regional varieties.

For /r/ coda, Brazilian accents are divided into three main classes. The \textit{tap} class corresponds to the alveolar flap \textipa{[R]}, found in places like São Paulo city and Porto Alegre, for example. The \textit{carioca} class corresponds to the velar or glottal fricative [\textipa{x}]/[\textipa{h}], characteristic of Rio de Janeiro and the surrounding Fluminense dialect region. The \textit{caipira} class corresponds to the retroflex vibrant \textipa{[\textrtailr]}, a realization associated with the rural interior of São Paulo and the state of Paraná.

For /d/--/t/ palatalization, a binary distinction is defined. The \textit{palatalized} class covers speakers who categorically affricate the coronal stops before the phonological vowel /i/, producing \textipa{[dZ\super{i}]} and \textipa{[tS\super{i}]} respectively, a pattern that is dominant across Rio and São Paulo. The \textit{non-palatalized} class covers speakers who preserve the plain stop \textipa{[d\super{i}]} and \textipa{[t\super{i}]}, a realization more prevalent in northeastern and some southern regional varieties.

The human label annotation was done through a web interface built specifically for this task, where ten samples (or more if needed) for each speaker were shown and the authors could choose the labels in terms of the classes described above.
 
\section{Accent markers classification}

Speaker-level classification is evaluated for each of the three phonological tasks using a stratified grouped cross-validation protocol. To remove any confounding effect of class imbalance, speakers are undersampled to equalize the number of speakers per class before evaluation; test folds contain five speakers and are stratified so that each class is represented in every fold. Also, speakers are isolated between folds to prevent training and evaluation from being done with the same speaker. Performance is reported as speaker accuracy (majority vote over all clips extracted from that speaker), using multiple classifiers: XGBoost, Logistic Regression, and linear SVM.
\subsection{Feature choices}

Table~\ref{tab:marker_results} reports the best speaker accuracy obtained across all feature and classifier configurations for each task, together with per-class recall.
\begin{table}[h]
\centering
\caption{Best speaker accuracy per phonological task under balanced stratified cross-validation.}
\label{tab:marker_results}
\setlength{\tabcolsep}{5pt}
\begin{tabular}{llccc}
\toprule
\textbf{Task} & \textbf{Classes} & \textbf{Best features} & \textbf{Spk. Acc.} & \textbf{Class recall} \\
\midrule
\multirow{2}{*}{/s/ coda}
  & \textit{chiado} \textipa{[S]}               & \multirow{2}{*}{ZIPA}          & \multirow{2}{*}{\makecell{1.00\\\scriptsize$\pm0.00$}} & 1.00 \\
  & \textit{sibilant} \textipa{[s]}             &                                &                                                         & 1.00 \\
\midrule
\multirow{3}{*}{/r/ coda}
  & \textit{carioca} [\textipa{x}]/[\textipa{h}]   & \multirow{3}{*}{Sp.+ZIPA+Allo} & \multirow{3}{*}{\makecell{0.85\\\scriptsize$\pm0.22$}} & 0.94 \\
  & \textit{tap} \textipa{[R]}                  &                                &                                                         & 0.71 \\
  & \textit{caipira} \textipa{[\textrtailr]}           &                                &                                                         & 0.88 \\
\midrule
\multirow{2}{*}{/d/--/t/}
  & \textit{palat.} \textipa{[dZ\super{i}]}     & \multirow{2}{*}{Sp.+ZIPA}      & \multirow{2}{*}{\makecell{0.88\\\scriptsize$\pm0.14$}} & 0.90 \\
  & \textit{non-palat.} \textipa{[d\super{i}]}  &                                &                                                         & 0.80 \\
\bottomrule
\end{tabular}
\end{table}

The dataset for evaluation contain 40 manually annotated speakers for the /s/-coda task (20 per class), 51 speakers for the /r/-coda task (17 per class), and 20 speakers for /d/-/t/-palatalization (10 per class). Winner features for /s/-coda were ZIPA alone, with perfect speaker accuracy in the experiment for both classes. For /r/-coda, the combination of spectral features with ZIPA and Allossaurus logits beat other competitors with 0.85 accuracy over speakers, and tap /r/ is, in general, the hardest class. For /d/-/t/ palatalization, spectral features with ZIPA got the best results among the lightweight models. The full ablation study for choosing the models is available at the companion webpage\footnote{https://gpa-smt-ufrj.github.io/accent-features}.

\subsection{Window choice} To select the best temporal window for spectral and MFCC features, we perform a grid search over window size $w \in \{40,80,120,160\}$\,ms and center offset $\delta \in \{-40,-20,0,+20,+40\}$\,ms relative to the ZIPA-detected spike, restricting to valid pairs where $|\delta|+w/2 \leq 80$\,ms. Sub-windows are applied at feature computation time over pre-extracted 160\,ms clips. Table~\ref{tab:window_best} reports the optimal window configuration per feature and task, also evaluated under a balanced cross validation protocol. 

\begin{table}[ht]
\centering
\caption{Optimal window size $w$ and offset $\delta$ per feature and task.}
\label{tab:window_best}
\setlength{\tabcolsep}{4pt}
\begin{tabular}{lcccccc}
\toprule
& \multicolumn{2}{c}{\textbf{/s/ coda}} & \multicolumn{2}{c}{\textbf{/r/ coda}} & \multicolumn{2}{c}{\textbf{/d/--/t/ palat.}} \\
\cmidrule(lr){2-3}\cmidrule(lr){4-5}\cmidrule(lr){6-7}
\textbf{Feature} & $w$ & $\delta$ & $w$ & $\delta$ & $w$ & $\delta$ \\
\midrule
MFCC & 120\,ms & $-$20\,ms & 120\,ms & 0\,ms & 160\,ms & 0\,ms \\
Spec.  & 40\,ms  & $+$20\,ms & 40\,ms  & $+$40\,ms & 80\,ms & $+$20\,ms \\
\bottomrule
\end{tabular}
\end{table}

For /s/ coda, MFCC benefits from a slight pre-onset bias ($\delta=-20$\,ms), capturing frication onset, while spectral features peak at a narrow post-onset window ($w=40$\,ms, $\delta=+20$\,ms), after which wider windows dilute the burst. For /r/ coda, MFCC peaks at a centered medium window ($w=120$\,ms, $\delta=0$), while Spec. peaks with a narrow post-onset configuration ($w=40$\,ms, $\delta=+40$\,ms), suggesting rhotic cues are concentrated in a short post-burst transition. For /d/-/t/ palatalization, MFCC improves at the full $w=160$\,ms window and Spec. has the largest length, which points out to the phenomena affecting the long consonant-vowel transition.

\subsection{Localization importance}

Accessing how feature extraction alignment can improve accent marker detection, another experiment is conducted. As a whole-utterance
baseline, SoTA deep learning models extract vectors from the entire recording and mean-pools
all frames without ZIPA localization. Then, the results for each marker detection task are compared with the best aligned models found in the stratified cross validation experiment.
\begin{table}[ht]
  \centering
  \caption{Speaker accuracy of whole-utterance embeddings vs.\ proposed aligned features.}
  \label{tab:gating_ablation}
  \setlength{\tabcolsep}{3pt}
  \renewcommand{\arraystretch}{0.95}
  \resizebox{\columnwidth}{!}{%
  \begin{tabular}{lccc}
  \toprule
  \textbf{Model}
    & \textbf{/s/ coda} & \textbf{/r/ coda} & \textbf{/d/-/t/} \\
  \midrule
  HuBERT          & $0.82\pm0.14$ & $0.42\pm0.17$ & $0.60\pm0.11$ \\
  Wav2Vec 2.0     & $0.85\pm0.18$ & $0.45\pm0.15$ & $0.73\pm0.18$ \\
  XLS-R        & $0.79\pm0.16$ & $0.45\pm0.18$ & $0.52\pm0.15$ \\
  XLSR-PT         & $0.74\pm0.24$ & $0.53\pm0.18$ & $0.69\pm0.12$ \\
  ECAPA-TDNN       & $0.69\pm0.25$ & $0.57\pm0.20$ & $0.50\pm0.12$ \\
  Resemblyzer     & $0.68\pm0.17$ & $0.40\pm0.27$ & $0.71\pm0.22$ \\
  \midrule
  \textbf{Our aligned best} & $\mathbf{1.00\pm0.00}$ & $\mathbf{0.85\pm0.22}$ & $\mathbf{0.88\pm0.14}$ \\
  \bottomrule
  \end{tabular}%
  }
\end{table}

Results on Table~\ref{tab:gating_ablation} shows that general-purpose models suffer from information dilution when processing full utterances, as using the extraction windows drastically improves accuracy across all phonetic tasks against the whole utterance baselines.

\begin{table*}[!t]
  \centering
  \caption{4-class Brazilian accent classification: per-class and macro F1
  (mean $\pm$ std over 30 repeated balanced CV runs,
  logistic regression). Best per column in \textbf{bold}.}
  \label{tab:4class-ne-f1}
  \begin{tabular}{lccccr}
  \toprule
  \textbf{Model} & \textbf{NE} & \textbf{RJ} & \textbf{SP} & \textbf{MG} & \textbf{Macro F1} \\
  \midrule
  ZIPA v2 (7D)       & $0.84 \pm 0.05$ & $\mathbf{0.84 \pm 0.03}$ & $\mathbf{0.81 \pm 0.03}$ & $0.81 \pm 0.05$ & $\mathbf{0.83 \pm
  0.02}$ \\
  \midrule
  HuBERT (768D)      & $\mathbf{0.95 \pm 0.02}$ & $0.78 \pm 0.04$ & $0.72 \pm 0.04$ & $\mathbf{0.92 \pm 0.03}$ & $\mathbf{0.84 \pm
  0.02}$ \\
  XLSR-PT (1024D)    & $\mathbf{0.95 \pm 0.03}$          & $0.75 \pm 0.03$ & $0.72 \pm 0.03$ & $\mathbf{0.92 \pm 0.04}$ & $\mathbf{0.84 \pm
  0.02}$ \\
  XLS-R (1024D)      & $\mathbf{0.95 \pm 0.02}$          & $0.73 \pm 0.04$ & $0.64 \pm 0.04$ & $\mathbf{0.93 \pm 0.03}$ & $0.81 \pm
  0.02$ \\
  Wav2Vec 2.0 (768D)    & $0.86 \pm 0.04$          & $0.77 \pm 0.03$ & $0.67 \pm 0.04$ & $0.90 \pm 0.03$ & $0.80 \pm 0.03$ \\
  Resemblyzer (256D) & $0.86 \pm 0.05$          & $0.68 \pm 0.04$ & $0.66 \pm 0.05$ & $0.89 \pm 0.06$ & $0.77 \pm 0.03$ \\
  ECAPA-TDNN (192D)  & $0.75 \pm 0.06$          & $0.61 \pm 0.04$ & $0.59 \pm 0.05$ & $0.85 \pm 0.04$ & $0.70 \pm 0.03$ \\
  \bottomrule 
  \end{tabular}
\end{table*}
\section{Accent detection}
After analyzing the effectiveness of the features in classifying phoneme markers, two experiments were conducted applying the best classifiers to the accent detection task. 

Notably, in Brazilian Portuguese, the datasets available for this purpose rely on sociolinguistic labels, which, as discussed before, might induce bias or misclassifications. However, the main motivation for building marker classifiers is to detect and classify accents; therefore, the evaluation is performed on those datasets to maintain a fair comparison with the established baselines,  but with specific precautions to avoid training leakage and attempt to reduce bias.

Following the recent literature on Brazilian Portuguese accent detection, the first experiment consisted in evaluating the performance of classifiers trained for the marker detection task in a cross-dataset accent detection scenario.
To do this, we replicated the experiment setup described in~\cite{matos2023accent}, and compared the proposed system against that baseline. Results for this experiment are shown in Table~\ref{tab:pe_sp_crossdataset}.
\begin{table}[h]
\centering
\caption{Cross-dataset binary PE vs SP accent classification
         (train: Spotify-B; test: CORAA-ASR).
         P = precision, R = recall, F1 = F1-score (\%).}
\label{tab:pe_sp_crossdataset}
\setlength{\tabcolsep}{5pt}
\begin{tabular}{lccccccc}
\toprule
& \multicolumn{3}{c}{\textbf{PE}} & \multicolumn{3}{c}{\textbf{SP}}
& \multirow{2}{*}{\textbf{Overall}} \\
\cmidrule(lr){2-4}\cmidrule(lr){5-7}
\textbf{System} & P & R & F1 & P & R & F1 & \\
\midrule
CNN-LSTM \cite{matos2023accent}
  & 50\% & 96\% & 66\% & 73\% & 11\% & 19\% & 43\% \\
XLSR-53 ft.\ \cite{matos2023accent}
  & 68\% & 99\% & 80\% & \textbf{99\%} & 55\% & 70\% & 75\% \\
\midrule
\textbf{Ours (7D)}
  & \textbf{88\%} & \textbf{99\%} & \textbf{93\%}
  & 93\% & \textbf{57\%} & \textbf{70\%}
  & \textbf{82\%} \\
\bottomrule
\end{tabular}
\end{table}

Using a simple 7-dimensional vector containing the mean output probabilities for each class (2 for /s/-coda, 3 for /r/-coda, and 2 for /d/-/t/ palatalization) over each speaker and a logistic regression algorithm, it was possible to achieve results superior to those of the established baseline. Also, in Table~\ref{tab:logreg_coef}, interpretability results are shown. Carioca /r/-coda presence points towards PE, where other /r/-coda classes contribute to SP, and this task contributes the most for the classification. The /s/-coda ``chiado'' and non-palatalization of /d/-/t/ also point to PE. All these coefficients align with the regional accent realization data distribution presented in the ALiB~\cite{cardoso2014atlas} books.
\begin{table}[!h]
\centering
\caption{Logistic regression coefficients for binary PE vs SP classification}
\label{tab:logreg_coef}
\begin{tabular}{llrl}
\toprule
\textbf{Task} & \textbf{Phoneme class} & \textbf{Coeff.} & \textbf{Direction} \\
\midrule
/r/-coda     & carioca [\textipa{x}]/[\textipa{h}]      & $-$0.42 & $\to$ PE \\
/r/-coda     & tap [\textipa{R}]             & $+$0.36 & $\to$ SP \\
/r/-coda     & caipira \textipa{[\textrtailr]}       & $+$0.36 & $\to$ SP \\
/s/-coda     & chiado [\textipa{S}]          & $-$0.24 & $\to$ PE \\
/s/-coda     & sibilant [s]                  & $+$0.20 & $\to$ SP \\
/d/-/t/ palat. & non-palatalized               & $-$0.05 & $\to$ PE  \\
/d/-/t/ palat. & palatalized [\textipa{tS/dZ}] & $+$0.05 & $\to$ SP  \\
\bottomrule
\end{tabular}

\end{table}

The second experiment is a broader accent detection setup. Gathering data from multiple datasets, 4 accent classes were built using data from two or more sources for each region, to avoid recording and environmental shortcuts. The classes are: Norhteast (NE), with 135 speakers coming from ColingPB and TAGARELA\footnote{https://huggingface.co/datasets/freds0/TAGARELA}; carioca (RJ), 27 speakers from CETUC~\cite{alencar2008speech} and Sotaque Brasileiro\footnote{https://sotaque-brasileiro.github.io/}; paulista (SP), with 24 speakers from Sotaque Brasileiro and Ynoguti~\cite{yneguti1999}; and mineiro (MG, Belo Horizonte region), with 89 speakers taken from C-ORAL Brasil as distributed in CORAA~\cite{candido2023coraa}, Sotaque Brasileiro and Ynoguti. It should be noted that this arbitrary class definition was established to use only publicly available pt-BR data and guarantee the multiple sources per region. The idea of the experiment is not to build a general or broad-sense accent detection classifier, but rather to measure the performance of the proposed marker detectors compared to large pretrained models. 

In this sense, the procedure was to measure the accuracy of the 7-D vector of marker classification probabilities against deep learning baselines. The extracted baselines in this experiments are per-utterance means for each model. The measurements were done through balanced stratified cross validation with 24 speakers per class at a time (with 30 different sampling repetitions), and speaker isolation through folds. Results are shown in Table~\ref{tab:4class-ne-f1}, and point out to a very similar overall accuracy (Macro F1-Score) between the proposed strategy and best SoTA SSL models.


\section{Conclusions and future works}
This work demonstrated that extracting low-dimensional features at accent marker instants enables accent classification accuracy competitive with massive deep learning models while remaining interpretable and computationally lightweight. 

By using ZIPA as a localization engine, three regional landmarks are isolated: /s/-coda, /r/-coda, and /d/–/t/ palatalization, showing that local spectral and phoneme-probability features extracted at those precise windows carry strong dialectal signal. With reasonable accent marker detection quality and relevant improvement against utterance-based classification, the employment of such localized features can surpass fine-tuned SSL baselines in cross-dataset accent detection scenarios with explainable results that align with the regional marker distributions documented in the literature. 

It is important to note, however, that even though the accent markers employed in this work can disambiguate between many Brazilian accents, they are not powerful enough for general pt-BR accent detection. Many variations in the language will rely on other markers such as /l/ palatalization or /e/ and /o/ vowel openness, as found in ALiB maps. Also, prosodic tendencies may be the predominant difference between languages spoken in certain regions. For these reasons, a broader analysis is needed to address more markers and to take prosody lines into account.

Nevertheless, this work opens a new paradigm for accent exploration in pt-BR, showing that sociolinguistic labels are not mandatory for building useful accent classifiers for the country's linguistic variations. It also shows the possibility of avoiding the need for large data and training pipelines with the employment of lightweight signal processing techniques and small model adaptations.

\section*{Aknowledgement}
This work was submitted to the XLIV Brazilian Symposium on Telecommunications and Signal Processing  (SBrT 2026).

\bibliographystyle{ieeetr}
\bibliography{references}

\end{document}